\documentclass[aps,prx,twocolumn,groupedaddress,superscriptaddress,amsfonts,amssymb,amsmath,citeautoscript,longbibliography,a4paper,floatfix,nofootinbib]{revtex4-1}

\usepackage[utf8]{inputenc}
\usepackage[english]{babel}

\usepackage{microtype} 
\usepackage{xspace} 

\usepackage{txfonts}  
\usepackage{txfontsb} 

\usepackage{bm} 

\usepackage{xcolor}
\usepackage[]{graphicx} 
\graphicspath{{figs/}}

\usepackage[]{booktabs}
\usepackage{array}
\usepackage{layouts}
\usepackage{multirow}

\usepackage{enumerate}
\usepackage[inline]{enumitem}

\usepackage{xr-hyper} 
\usepackage{hyperref}
\hypersetup{colorlinks,
	linkcolor={blue!75!black!80!yellow},
	citecolor={blue!75!black!80!yellow},
	urlcolor={blue!75!black!80!yellow},
	filecolor={blue!75!black!70!yellow}
}

\usepackage[capitalize,nameinlink]{cleveref}

\crefname{subequations}{Eqs.}{Eqs.} 
\Crefname{subequations}{Eqs.}{Eqs.}
\crefformat{subequations}{#2Eqs.~(#1)#3}
\Crefformat{subequations}{#2Eqs.~(#1)#3}
\crefname{page}{p.}{p.} 
\crefname{sitable}{Supp. Table}{Supp. Tables}
\crefname{sifigure}{Supp. Fig.}{Supp. Fig.}
\crefname{sisection}{Supp. Section}{Supp. Sections}

\usepackage{placeins}

\usepackage{siunitx}
\DeclareSIUnit[number-unit-product = ]\percent{\char`\%} 

\usepackage[centering,hmargin=16mm,tmargin=30mm,bmargin=26mm]{geometry}

\usepackage{soul}

\usepackage{textcomp} 
\usepackage{xifthen}
\usepackage{etoolbox}
\newboolean{togglecomments}
\newboolean{toggletodos}
\newboolean{togglechanges} 

\setboolean{togglecomments}{true}
\setboolean{toggletodos}{true}
\setboolean{togglechanges}{false}

\setboolean{togglechanges}{true}
\setboolean{toggletodos}{false}
\setboolean{togglecomments}{false}

\newcommand{\textblacksquare}{$\blacksquare$}
\newcommand{\todo}[1]{\ifbool{toggletodos}%
	{\textcolor{magenta!60!black}{\small\textsf{{}\textsuperscript{\textsc{\textsf{todo}}}}[#1]}} 
	{}}     
\newcommand{\comment}[2]{\ifbool{togglecomments}%
		{\textcolor{blue!70!black}{\small\sf\textsuperscript{\textsc{\textsf{#1}}}[#2]}} 
		{}}     
\newcommand{\help}[1]{\ifbool{togglecomments}%
	{\textcolor{red!60!black}{\small\textsf{{}\textsuperscript{\textsc{\textsf{help}}}}[#1]}} 
	{}}     
\newcommand{\swap}[2]{\ifbool{togglechanges}
	{#2}  
	{\textcolor{red!70!black}{[#1]}\textrightarrow{}\textcolor{green!50!black}{[#2]}}}
\newcommand{\remove}[1]{\ifbool{togglechanges}
	{}    
	{\textcolor{red!70!black}{#1}}}
\newcommand{\inset}[1]{\ifbool{togglechanges}
	{#1}  
	{\textcolor{green!50!black}{#1}}}
\newcommand{\citeremind}[1]{\ifbool{togglechanges}
    {}
    {%
	[\textcolor{blue!75!black!80!yellow}{\textblacksquare%
		\ifthenelse{\isempty{#1}}{}{\textsuperscript{\tiny\textsf{#1}}}%
	}]\xspace}
	}

\newcommand{\optional}[1]{\ifbool{togglechanges}
	{}
	{\textcolor{orange!70!gray}{#1}}}

\renewcommand{\paragraph}[1]{\vskip 1ex\noindent\textbf{#1.}~}

\usepackage[eulergreek]{sansmath}
\makeatletter
\renewcommand\@make@capt@title[2]{%
    \@ifx@empty\float@link{\@firstofone}{\expandafter\href\expandafter{\float@link}}%
    \sisetup{math-sf=\textsf}%
    \sansmath\sffamily\textbf{#1\@caption@fignum@sep}#2 
}%

\makeatother

\newcommand{\appropto}{\mathrel{\vcenter{
			\offinterlineskip\halign{\hfil$##$\cr
				\propto\cr\noalign{\kern.2pt}\sim\cr\noalign{\kern-2.5pt}}}}}

\makeatletter
\newcommand{\raisemath}[1]{\mathpalette{\raisem@th{#1}}}
\newcommand{\raisem@th}[3]{\raisebox{#1}{$#2#3$}}
\makeatother

\DeclareFontFamily{U}{mathx}{\hyphenchar\font45}
\DeclareFontShape{U}{mathx}{m}{n}{<5> <6> <7> <8> <9> <10>
                                  <10.95> <12> <14.4> <17.28> <20.74> <24.88>
                                  mathx10}{}
\DeclareSymbolFont{mathx}{U}{mathx}{m}{n}
\DeclareFontSubstitution{U}{mathx}{m}{n}
\DeclareMathAccent{\widebar}{0}{mathx}{"73}

\usepackage{filecontents}
\makeatletter	
\newcommand*{\addFileDependency}[1]{
  \typeout{(#1)}	
  \@addtofilelist{#1}	
  \IfFileExists{#1}{}{\typeout{No file #1.}}	
}	
\makeatother

\usepackage{float}
\makeatletter
\let\newfloat\newfloat@ltx
\makeatother
\usepackage{algorithm}
\usepackage{algpseudocode}
\makeatletter
\renewcommand{\fname@algorithm}{Algorithm }
\makeatother
\DeclareMathOperator*{\argmin}{arg\,min}
\DeclareMathOperator*{\argmax}{arg\,max}

\usepackage{float}
\begin{document}

\title{Active learning for photonic crystals}

\def\mitaffil{Department of Physics, Massachusetts Institute of Technology, Cambridge, Massachusetts, USA}
\def\miteecsaffil{Department of Electrical Engineering and Computer Science, Massachusetts Institute of Technology, Cambridge, Massachusetts, USA}
\author{Ryan Lopez}
\email{rnlopez@mit.edu}
\affiliation{\mitaffil}
\author{Charlotte Loh}
\affiliation{\miteecsaffil}
\author{Rumen Dangovski}
\affiliation{\miteecsaffil}
\author{Marin~Solja\v{c}i\'{c}}
\affiliation{\mitaffil}

\begin{abstract}
Active learning for photonic crystals explores the integration of analytic approximate Bayesian last layer neural networks (LL-BNNs) with uncertainty-driven sample selection to accelerate photonic band gap prediction. We employ an analytic LL-BNN formulation, corresponding to the infinite Monte Carlo sample limit, to obtain uncertainty estimates that are strongly correlated with the true predictive error on unlabeled candidate structures. These uncertainty scores drive an active learning strategy that prioritizes the most informative simulations during training. Applied to the task of predicting band gap sizes in two-dimensional, two-tone photonic crystals, our approach achieves up to a 2.7× reduction on average in required training data compared to a random sampling baseline while maintaining predictive accuracy. The efficiency gains arise from concentrating computational resources on high uncertainty regions of the design space rather than sampling uniformly. Given the substantial cost of full band structure simulations, especially in three dimensions, this data efficiency enables rapid and scalable surrogate modeling. Our results suggest that analytic LL-BNN based active learning can substantially accelerate topological optimization and inverse design workflows for photonic crystals, and more broadly, offers a general framework for data efficient regression across scientific machine learning domains.
\end{abstract}
\maketitle

\section{Introduction}
 Modern scientific and engineering workflows increasingly depend on machine‑learning surrogates to circumvent the steep computational or experimental costs of high‑fidelity simulations. For instance, in photonics, computing the full band structure of a photonic crystal can be computationally expensive, severely limiting the exploration of large design spaces and slowing inverse‑design cycles \cite{Loh2022Surrogate}. Data‑efficient strategies that strategically select only the most informative simulations are therefore essential to accelerate discovery and optimization in such domains.

Active learning (AL) addresses this challenge by iteratively querying labels for unlabeled points deemed most informative by a model \cite{settles2009active}. Early pool‑based AL in classification relied on uncertainty sampling, selecting points near decision boundaries via margin or entropy measures, to achieve significant label savings on image and text datasets \cite{lewis1995sequential, tong2001support}. Batch active learning extends the standard AL setting by querying multiple points simultaneously rather than a single example per iteration. Supporting larger batch sizes is important in modern training pipelines because model updates are expensive, but naïvely selecting the top-$k$ uncertain points often yields redundant and highly correlated samples \cite{4798162}. To overcome this problem, recent batch AL methods have integrated diversity constraints with uncertainty. For example, BatchBALD extended pointwise mutual information to batches through a greedy $(1-1/e)$ approximation using MC‑dropout BNNs \cite{ad3a75fa2a26e6f69b7059466828f49492d31789}, whereas BADGE derives pseudo-labels and applies $k$‑means++ clustering on gradient embeddings for joint uncertainty–diversity sampling \cite{cf5a21684aefb1b8db6e0490167636d245396095}. Batch Active Learning at Scale further combines margin based uncertainty with round-robin style sampling from clusters to support massive batch sizes (up to millions) with provable label‑complexity bounds \cite{998bc35af7c44a37f0eaca96c181b6aed9b7d489}, and principled determinantal point processes have been efficiently approximated with MCMC \cite{ebcf880fb4ffffd3140d97c226b7040b1a46bf73}. Scalability via posterior‑sparse approximations, akin to Bayesian coresets, has also been demonstrated for both classification and regression settings \cite{37660ead36b43de2e262e7471e19b32b45d39725}. Semi‑supervised hybrids further ameliorate computational costs by pretraining encoders on unlabeled data before fitting a stochastic prediction head \cite{407c94452ceeaf3e11121429c50fd75094094358}.

Despite these advances in classification, many scientific applications require continuous predictions rather than discrete labels. Pool‑based batch AL for regression has been explored through “black‑box” methods like B3AL, which uses kernelized embeddings of virtual ensembles to generalize white‑box acquisition functions to arbitrary models \cite{073c97a6bbebcebd0f9b5d441c194d397b6654cb}, and Gaussian‑process acquisitions such as maximum expected prediction error (MEPE) have been applied to machine‑learning force‑field training \cite{00bfc9643b0c7fa719c79bfa8dd330846792c426}. In chemical reactivity and potential‑energy surface sampling, committee‑variance sampling with message‑passing neural nets has shown dramatic efficiency gains by focusing on high-variance molecular configurations \cite{1ba98aaaae53a8ce7f1dec69f7d11319b1e0b43d}. Active learning has recently been applied to improve data efficiency in nanophotonic design problems, including surrogate modeling for PDE-based metasurface components and large-scale inverse design across diverse design spaces \cite{AL_PDE_metasurface, AL_nanophotonics_inverse_design}. However, these approaches focus on metasurface design rather than analytic uncertainty–driven regression for photonic-crystal band-gap prediction.

Approximate Bayesian neural networks (BNNs) offer principled uncertainty quantification, but conventional approaches rely on Monte Carlo sampling, often via MC-dropout or deep ensembles, with tens to hundreds of forward passes per candidate, which can be prohibitive in high-throughput settings \cite{da5c65b0ac8b525c3d3d4889bf44d8a48d254a07, 74784d758960f07bd6ef9131585f262970450f34}. Approximate last layer Bayesian neural networks (LL-BNNs), in contrast, permit a closed-form, infinite-MC-limit solution: both the predictive variance and the Kullback–Leibler (KL) regularizer admit analytic expressions, eliminating sampling noise and reducing acquisition overhead to a single forward pass plus a few matrix operations. Recent work on variational Bayesian last layers (VBLL) generalizes this formulation by introducing a deterministic variational objective that achieves sampling-free inference with only quadratic complexity in the last-layer width \cite{harrison2024variationalbayesianlayers}. Complementary directions such as evidential deep learning provide non-Bayesian yet efficient uncertainty estimation by directly predicting distributional evidence parameters, enabling well-calibrated uncertainty at nearly no additional inference cost and powering uncertainty-guided active learning in molecular property prediction \cite{doi:10.1021/acscentsci.1c00546}.

\begin{figure}[htbp]
\centering\includegraphics[width=7cm]{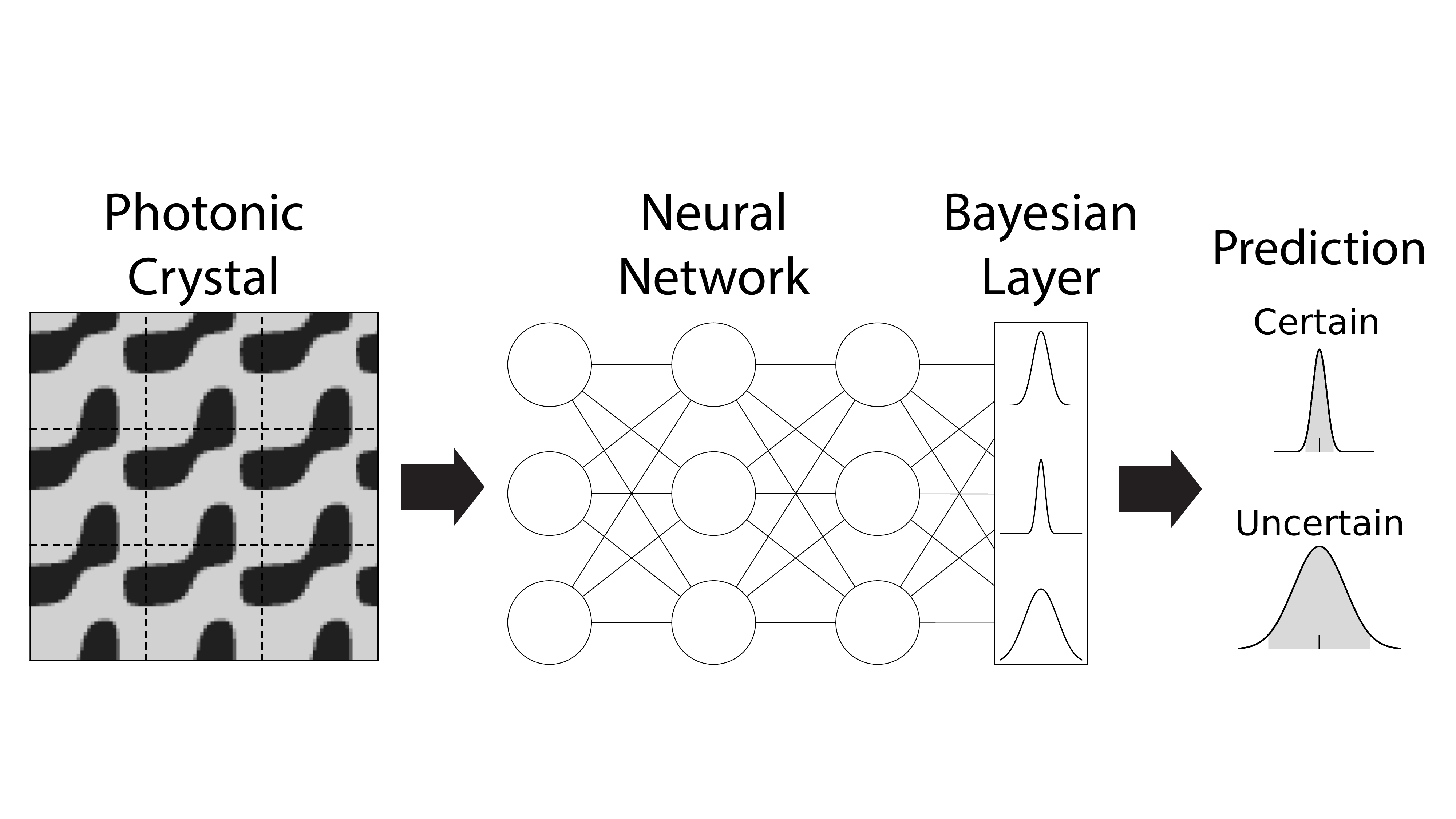}
\caption{\label{fig:bnn_overview} \textbf{Active‐Learning Pipeline for 2D Photonic‐Crystal Band‑Gap Prediction.} Starting from the dielectric‐constant map of a 2D two‑tone photonic‑crystal unit cell (augmented via symmetry operations), we feed each sample into a deep neural network whose final layer is approximate
Bayesian. In this last layer, each weight and bias is treated as a Gaussian random variable, so a single input yields a predictive distribution rather than a point estimate. When evaluating unlabeled candidates, we compute the output variance as an uncertainty score and select the highest‐variance samples for expensive band‑structure simulations. By iteratively adding only the most informative points, the model rapidly improves its accuracy with far fewer training examples.}
\end{figure}

In this work, we introduce an analytic LL-BNN active learning framework tailored to regression of two‑dimensional photonic crystal band gaps. We train the full network jointly, feature extractor and approximate Bayesian last layer, on an initial pool of 50 samples and iteratively select the next 50 most uncertain candidates by evaluating the closed form variance. On a dataset of 11,376 two‑tone 2D photonic crystals, our uncertainty‑driven sampling achieves up to 2.7× data savings on average compared to random selection. To our knowledge, this is the first application of analytic last‑layer approximate BNN active learning to photonic‑crystal regression, and it offers a practical, data‑efficient surrogate for photonic band gap estimations. A high-level overview of our active learning workflow is shown in Figure \ref{fig:bnn_overview}.

\section{\label{sec:level1}Methods}
\subsection{Approximate Bayesian last layer}\label{sec:bayesian-last-layer}

An approximate Bayesian last layer can be appended to any neural network, both trained jointly, to provide uncertainty estimates with minimal overhead.  Denote by $x\in\mathbb{R}^d$ the deterministic features from the backbone and $y\in\mathbb{R}$ the regression target. The last layer computes $W^\top x + b$ where the weight vector $W \in \mathbb{R}^d$ and bias $b \in \mathbb{R}$ are treated as random variables. Each component is distributed independently as
\begin{equation}
    W_i \sim \mathcal{N}(\mu_{w,i},\, \sigma_{w,i}^2), 
\qquad
b \sim \mathcal{N}(\mu_b,\, \sigma_b^2)
\end{equation}
where $\mu_w, \sigma_w \in \mathbb{R}^d$ and $\mu_b, \sigma_b \in \mathbb{R}$ are trainable parameters. These independent Gaussians together define the variational posterior $q(W,b)$. We then place a standard normal prior over the variables:
\begin{equation}
    p(W, b) = \mathcal{N} \left( \begin{bmatrix} W \\ b\end{bmatrix} ; 0, I\right)
\end{equation}
and train by minimizing a loss inspired by the negative evidence‑lower‑bound (ELBO):
\begin{equation}\label{eq:total-loss}
  \mathcal{L}
  = \underbrace{\mathbb{E}_{q(W,b)}\!\bigl[(y - W^\top x - b)^2\bigr]}_{\mathcal{L}_{\mathrm{pred}}}
  + \,
    \underbrace{D_{\mathrm{KL}}\bigl[q(W,b)\,\|\,p(W,b)\bigr]}_{\mathcal{L}_{\mathrm{KL}}}\,
\end{equation}
with $D_{\mathrm{KL}}(\cdot)$ denoting the Kullback-Leibler divergence. Since this is the only Bayesian layer, the predictive‑loss term admits an analytic closed form (the $\mathrm{MC}\!\to\!\infty$ limit of sampling):
\begin{align}
  \mathcal{L}_{\mathrm{pred}}
  &= \mathbb{E}_{q(W,b)}\bigl[(y - W^\top x - b)^2\bigr] \nonumber \\ 
  &= \bigl(y - \mu_w^\top x - \mu_b\bigr)^2
     + x^\top\Sigma_w\,x
     + \sigma_b^2
\end{align}
where $\Sigma_w=\operatorname{diag}(\sigma_{w,1}^2,\dots,\sigma_{w,d}^2)$.

Similarly, the KL term over independent Gaussians is
\begin{align}
  \mathcal{L}_{\mathrm{KL}}
  &= \sum_{i=1}^d
     D_{\mathrm{KL}}\bigl(
       \mathcal{N}(\mu_{w,i},\sigma_{w,i}^2)\,\|\,
       \mathcal{N}(0,1)\bigr) \nonumber\\ & \; \qquad +  D_{\mathrm{KL}}\bigl(
       \mathcal{N}(\mu_b,\sigma_b^2)\,\|\,
       \mathcal{N}(0,1)\bigr)
  \\
  &= \sum_{i=1}^d
     \Bigl[
       -\ln\sigma_{w,i}
       + \tfrac{\sigma_{w,i}^2 + \mu_{w,i}^2 - 1}{2}
     \Bigr] + 
     \Bigl[
       -\ln\sigma_b
       + \tfrac{\sigma_b^2 + \mu_b^2 - 1}{2}
     \Bigr]
\end{align}

From \eqref{eq:total-loss}, large posterior variances $\{\sigma_{w,i}^2,\sigma_b^2\}$ incur both higher predictive variance and a larger KL penalty, so the model learns to keep uncertainty low where possible.  In particular, the per-input uncertainty score
\begin{equation}
  s(x) \;=\; \operatorname{Var}\bigl[W^\top x + b\bigr]
        \;=\; x^\top\Sigma_w\,x + \sigma_b^2
\end{equation}
is available in closed form and used for uncertainty‑sampling. Selecting the highest-variance point is equivalent to greedily minimizing the model’s parameter uncertainty; further details are provided in Appendix~\ref{sec:level2}.

\subsection{\label{sec:level2}Deep ensemble uncertainty}
A deep ensemble estimates predictive uncertainty by training $N$ deterministic networks independently on the same labeled set, each initialized with different random weights. Deep ensembles are widely regarded as a strong baseline for uncertainty estimation \cite{lakshminarayanan2017simple} and serve here as a benchmark to contextualize the accuracy and uncertainty quality of the analytic LL-BNN. Given an input $x$, each network produces a scalar prediction $\hat{y}_n(x)$ for $n=1, \dots, N$. The ensemble prediction is taken as the mean output,
\begin{equation}
    \bar{y}(x) = \frac{1}{N} \sum_{n=1}^N \hat{y}_n(x)
\end{equation}
and the uncertainty score is the variance across predictions,
\begin{equation}
    s(x) = \frac{1}{N-1} \sum_{n=1}
^N \left(\hat{y}_n(x) - \bar{y}(x) \right)^2\end{equation}
This variance serves as the uncertainty score driving sample selection, with the key distinction from LL-BNNs that it requires training $N$ independent models per active learning iteration rather than one, increasing computational costs.

\subsection{\label{sec:level2}Uncertainty active learning}
We use Algorithm \ref{al_algorithm} to intelligently select samples for training. Uncertainty sampling typically outperforms random sampling because it actively selects the examples on which the current model is least confident, thereby maximizing information gain per label. By focusing on high‑uncertainty points, often near poorly explored regions of the input space, the model learns from the most ambiguous or informative data first, which accelerates error reduction and improves data efficiency. In contrast, random sampling treats all unlabeled points equally, wasting valuable labeling budget on regions the model already understands well. As a result, uncertainty sampling can achieve a given level of accuracy with far fewer labeled examples, making it especially advantageous when labels are costly or slow to obtain.

\begin{algorithm}[h]
\caption{Active learning for regression}
\begin{algorithmic}[1]
\State \textbf{Input:} Initial labeled set $L$, unlabeled set $U$, budget $B$, query batch size $q$
\While{$B > 0$}
    \State Train model $M$ on $L$
    \For{each sample $x \in U$}
        \State Compute uncertainty score $s(x)$ using $M$
    \EndFor
    \State Select a subset $S \subset U$ of $q$ samples with highest $s(x)$
    \State Query oracle to obtain true labels for $S$
    \State Update labeled set: $L \gets L \cup S$
    \State Update unlabeled pool: $U \gets U \setminus S$
    \State Update budget: $B \gets B - q$
\EndWhile
\State \textbf{Output:} Final model $M$ trained on $L$
\end{algorithmic}
\label{al_algorithm}
\end{algorithm}

\subsection{\label{sec:level2}Neural network architecture and training}
For our prediction neural network we use a deterministic feature-extraction network followed by an analytic approximate Bayesian layer. The feature network takes as input a single-channel $32 \times 32$ discretized permittivity map representing a two-dimensional photonic crystal unit cell. The feature extractor consists of two convolutional layers with 32 and 64 channels, respectively, each using $3 \times 3$ kernels and ReLU activations. The second convolutional layer is followed by $2 \times 2$ max pooling. The resulting feature maps are flattened and passed through a fully connected hidden layer with 128 output units and ReLU activation. This is then passed to the final approximate Bayesian layer that maps the 128-dimensional feature vector to a distribution of scalar outputs. For the deep ensemble, we train $N=10$ independent networks with the same architecture, except that the final layer is a standard deterministic linear layer with no variance parameters.

Training is performed within an active learning loop. The model is initialized using an initial labeled set of 50 samples and retrained from scratch at each iteration as new samples are acquired. At each iteration, 50 unlabeled samples are selected based on predictive uncertainty and added to the training set. Mini-batches of size 64 are used, and models are trained for 300 epochs per iteration using the Adadelta optimizer with an initial learning rate of 0.1. A cosine annealing learning-rate schedule is applied over the full training horizon.

\section{\label{sec:level1}Dataset}
We use the same 2D photonic crystal dataset generated in Loh et al.\cite{Loh2022Surrogate}; further details can be found in that work. Each 2D square PhC unit cell is parameterized by taking the level set of a Fourier‐sum function, summing nine plane waves with random complex coefficients, and thresholding it to produce a two‑tone permittivity profile $\epsilon(r) \in \{\epsilon_1, \epsilon_2\}$ with $\epsilon_i$ sampled uniformly in $[1,20]$. This yields distinct $32 \times 32$ pixel unit‑cell images. For each selected cell, the transverse‐magnetic band structure is computed using MIT Photonic-Bands, which employs the plane wave expansion method, with a $25 \times 25$ plane‑wave basis and a $25 \times 25$ $k$‑point grid, from which the largest band gap in the first 10 bands is directly extracted and used as the scalar regression label for training.

\begin{figure}[htbp]
\centering\includegraphics[width=7cm]{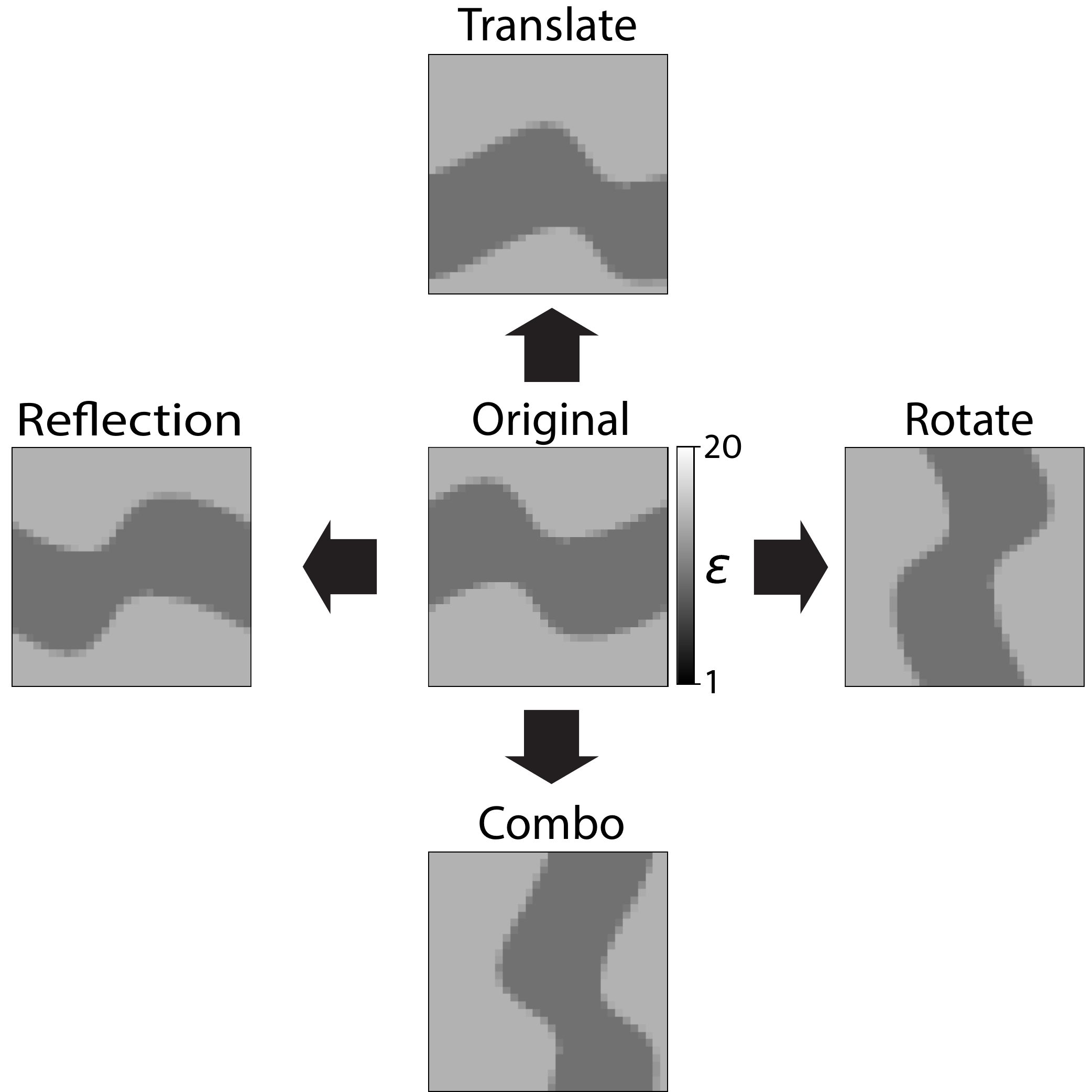}
\caption{\label{fig:epsart} \textbf{Symmetry-preserving data augmentation for 2D photonic crystal unit cells.} Rotations, reflections, and translations of the $32 \times 32$ permittivity maps leave the photonic band structure unchanged. During training, random combinations of these transformations are applied to each sample, effectively enlarging the dataset while enforcing physical invariance of the band gap labels.}
\end{figure}

To further enhance data efficiency, we exploit the intrinsic symmetries of the square photonic-crystal lattice, also used in Loh et al.\cite{Loh2022Surrogate}. Because the electromagnetic band structure is invariant under rigid motions that preserve the geometry—specifically, translations, rotations by 90°, 180°, and 270°, as well as mirror reflections across the horizontal and vertical axes—we use these operations as physics-preserving data augmentations (see Figure \ref{fig:epsart}). This ensures that the network learns features that are robust to lattice orientation and spatial placement, while effectively expanding the diversity of the training distribution without altering the underlying band-gap labels. Such augmentation improves generalization and accelerates convergence by enforcing the physical invariances of the system directly through the data. The dataset is split into disjoint training and test sets prior to active learning, with augmentation applied only to training samples after the split to prevent symmetry-induced leakage.

\section{Results}
\subsection{\label{sec:level2}Uncertainty estimation}
\begin{figure}[htbp]
\centering\includegraphics[width=7cm]{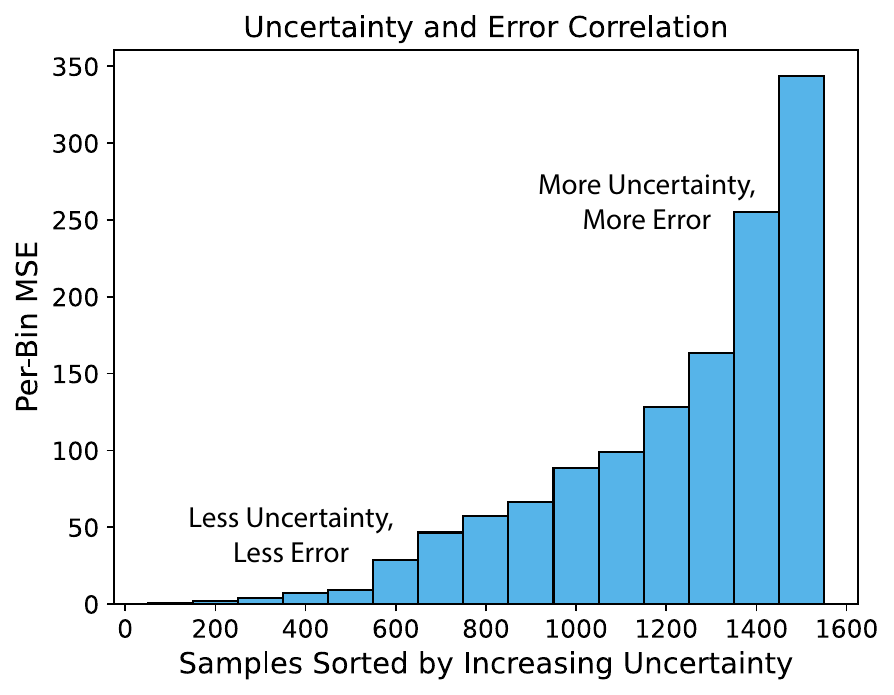}
\caption{\label{correlation_plot} \textbf{BNN Uncertainty Calibration.} After training on 50 randomly selected samples, we evaluate our approximate Bayesian last layer on the held‑out test set. For each test point, we compute the predictive variance $s(x)$ and the squared error $(y - \hat{y})^2$. We then sort all test samples by increasing $s(x)$, partition them into 15 bins of 100 samples each, and plot the mean squared error (MSE) of each bin along the sorted sample index. The clear upward trend shows that higher estimated uncertainty tends to correspond to larger true errors, 
suggesting a consistent monotonic relationship even when the overall model accuracy is low.}
\end{figure}

\begin{figure}[htbp]
\centering\includegraphics[width=7cm]{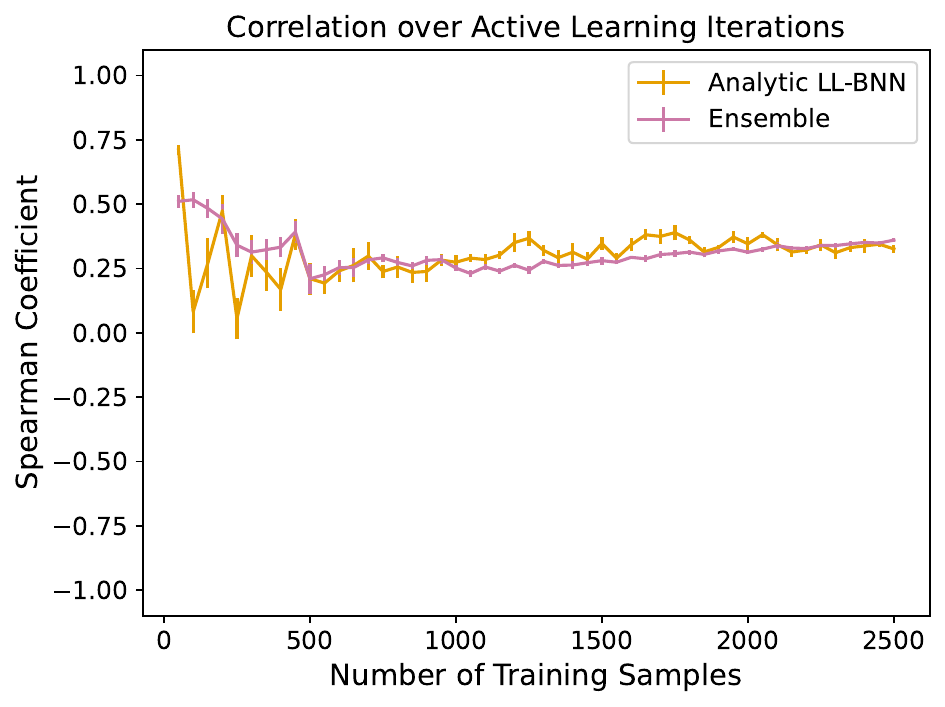}
\caption{\label{spearman_coeff_over_training} \textbf{Spearman Coefficient over Active Learning.} At each active learning step we compute the Spearman rank correlation between per-sample uncertainty scores $s(x)$ and per-sample squared errors $(y - \hat{y})^2$ on the test dataset. Both methods maintain a consistently positive coefficient across the full range of training dataset sizes, confirming that uncertainty estimates reliably rank samples in order of true error throughout training. The analytic LL-BNN and deep ensemble achieve similar Spearman coefficients, indicating comparable uncertainty quality despite the ensemble requiring approximately $N$ times more compute.}
\end{figure}

Our active learning strategy hinges on selecting samples with the highest model uncertainty, so it is only necessary that our approximate Bayesian last layer produces uncertainty estimates that rank points in order of true error, even if those estimates are not numerically perfectly calibrated. To verify this monotonicity, we proceed as follows:
\begin{enumerate}
    \item \textbf{Predictive statistics.}
    For each test input \(x\), record the model's predictive variance \(s(x)\) and its squared error \((y - \hat y)^2\).
    \item \textbf{Sorting and visualization.}
    Sort all test samples by increasing \(s(x)\). For visualization in Fig \ref{correlation_plot}, partition the sorted samples into bins of 100 samples each and compute the mean squared error within each bin \(\frac1{|\mathcal{B}|}\sum_{i\in\mathcal{B}}(y_i - \hat y_i)^2\).
  \item \textbf{Monotonicity metric.}
  Compute the Spearman rank correlation between per-sample uncertainty scores $s(x)$ and per-sample squared errors $(y-\hat{y})^2$ to quantify their monotonic relationship across the test set.
\end{enumerate}
Figure \ref{correlation_plot} displays one such calibration curve: bins with higher predictive variance exhibit proportionally higher true mean squared error (MSE). This suggests, even with a modestly performing model, that our uncertainty estimates tend to identify the samples on which the model is most likely to err, indicating that uncertainty‑driven sampling is likely to outperform random selection. Further per-sample visualization is provided in Appendix \ref{sec:level3}. Spearman's rank correlation coefficient, which ranges from +1 for a perfect direct monotonic relationship to -1 for a perfect inverse monotonic relationship, is a natural measure of this ranking quality as it directly quantifies whether higher predicted uncertainties correspond to larger true errors. The Spearman values throughout all active learning iterations are plotted in Figure \ref{spearman_coeff_over_training} for both the analytic LL-BNN and the deep ensemble, confirming that this positive monotonic relationship holds consistently across the full training pipeline. Notably, both methods achieve similar Spearman coefficients, suggesting that the analytic LL-BNN produces uncertainty estimates of comparable quality to the ensemble despite requiring only a single model.

\subsection{Performance curves}
\begin{figure}[htbp]
\centering\includegraphics[width=7cm]{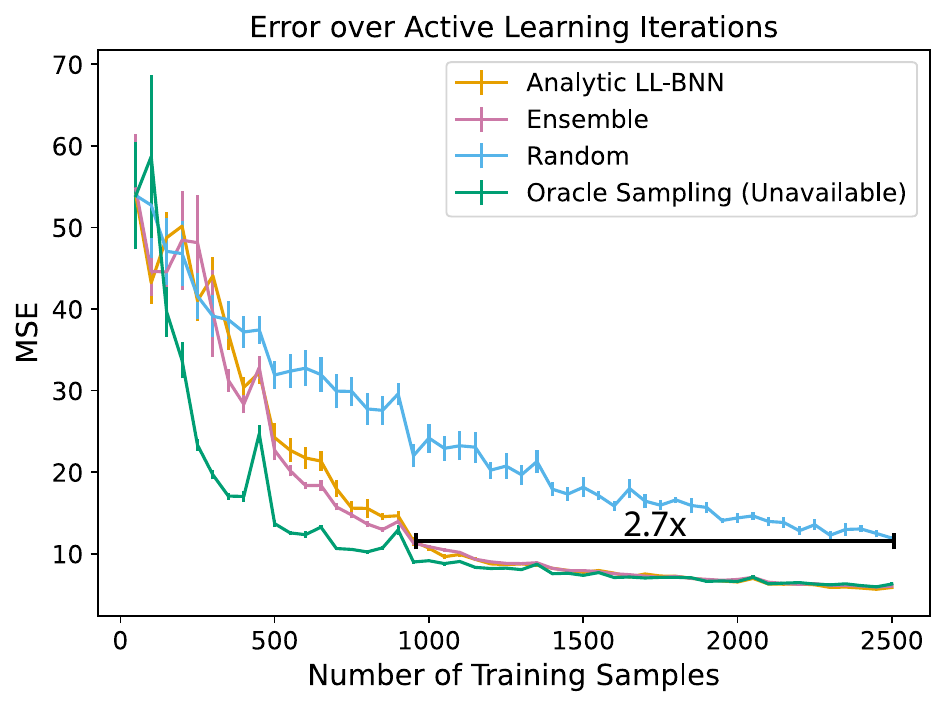}
\caption{\label{random_vs_uncertainty_phc} \textbf{Comparison of Random vs Uncertainty-Driven Sampling.} Test set MSE for photonic‐crystal band‑gap prediction as a function of total training samples, averaged over ten runs (each initialized with a different random pool of 50 examples). Blue shows standard random sampling; orange shows uncertainty sampling via our approximate Bayesian last layer; pink shows the deep ensemble. The green curve shows an "oracle" baseline that, at each iteration, greedily selects the 50 unlabeled samples with the highest true regression error—information unavailable in practice. This oracle is not globally optimal but provides an upper bound on what any greedy acquisition based on uncertainty could achieve. Error bars indicate standard error across runs. Both the analytic LL-BNN and the deep ensemble achieve lower error than random sampling at every budget level and reach the same accuracy as random sampling with roughly one-third the data. Notably, the two uncertainty-based methods perform comparably throughout training, while the analytic LL-BNN requires only a single model per iteration compared to $N$ for the ensemble.}
\end{figure}

We evaluate active learning on a dataset of 11,376 two‑tone, 2D photonic crystal unit cells, growing the training set from an initial pool of 50 up to 2,500 samples in increments of 50. At each iteration, we train the full network (including the Bayesian last layer), compute uncertainty scores for all unlabeled candidates, select the 50 samples with highest predictive variance, and retrain. Figure \ref{random_vs_uncertainty_phc} plots the resulting test set mean squared error versus cumulative training size, comparing uncertainty-driven acquisition using the analytic LL-BNN and deep ensemble to uniform random sampling (with matched random seeds and initial pools). The analytic LL-BNN and deep ensemble achieve comparable performance across training budgets, while the ensemble requires training multiple models per iteration.

Uncertainty sampling consistently outperforms random selection and reduces variability across runs. To quantify label efficiency, we define for each run a target MSE as the final accuracy achieved by random sampling at 2,500 training samples, and compute the number of labeled examples required by the analytic LL-BNN to reach this target. Averaged across all 10 runs, the analytic LL-BNN requires $2.70\times$ fewer labeled examples (95\% CI: [2.49, 2.91]) to match the final accuracy of random sampling, corresponding to a mean of 935 labeled examples (95\% CI: [870, 1000]) compared to 2,500 for random sampling. For a controlled comparison, the random sampling baseline uses the same network architecture and training protocol, differing only in the sampling strategy. The test set mean squared error is computed using the predictive mean of the model.

\section{\label{sec:level1}Discussion}

This study demonstrates that analytic approximate last-layer Bayesian neural networks (LL-BNNs) can serve as a practical engine for active learning in data-intensive photonics problems. By replacing Monte-Carlo sampling with a closed-form predictive variance, our approach
avoids sampling noise and yields stable, deterministic uncertainty estimates, enabling reliable large-scale screening of candidate structures with a single forward pass. In the context of two-dimensional, two-tone photonic crystals, we find that uncertainty-driven acquisition using the analytic LL-BNN reaches the same predictive accuracy as random sampling with roughly 2.7 times fewer labeled examples, and performs comparably to a deep ensemble while requiring substantially less compute. Beyond band gap prediction, the ability to append an analytic approximate Bayesian last layer to an arbitrary feature extractor suggests a scalable path to data-efficient regression across diverse scientific domains.

While our results are encouraging, several limitations remain. The present framework uses pure uncertainty sampling without explicit diversity constraints, which can lead to redundant selections when highly uncertain samples cluster in similar regions of the input space. Empirical quantification of batch redundancy via feature-space pairwise distances is provided in Appendix \ref{sec:level4}, suggesting that batches do not collapse to near-duplicate selections in practice. Incorporating batch-diverse acquisition functions such as determinantal point processes or BatchBALD could further improve performance in active-learning iterations. In addition, although our uncertainty estimates rank samples accurately by true error, absolute calibration was not the focus of this study. For safety-critical applications or multi-objective optimization, improved calibration (e.g., via evidential deep learning \cite{amini2020deep} or temperature scaling \cite{guo2017calibration}) may become important.

Future work could explore integrating this analytic LL-BNN active learning into full inverse-design loops for photonic crystals, where a surrogate model is repeatedly queried inside an optimizer. Because the Bayesian component is restricted to the last layer, the approach is compatible with more expressive feature extractors such as graph neural networks, transformers, or physics-informed architectures. Another promising direction is applying this approach to photonic crystals with defects, where simulations are substantially more computationally expensive and data efficiency becomes increasingly important. The same framework could be applied using supercell permittivity maps as inputs and defect band-structure simulations as labels, with uncertainty-driven acquisition selecting the most informative defect configurations. Additionally, hybridizing uncertainty sampling with semi-supervised pretraining or active data augmentation could further reduce the initial labeled pool. Finally, benchmarking against Monte-Carlo-based BNNs and Gaussian-process acquisitions on three-dimensional photonic-crystal datasets would clarify the trade-offs between analytic variance, diversity, and scalability at larger problem sizes.

Although this work focuses on scalar band gap prediction, the proposed active learning framework extends naturally to vector-valued outputs such as full photonic band structures. In that setting, the model would predict band frequencies across sampled $k$-points, and uncertainty could be defined by aggregating predictive variance across bands and $k$-points. This would enable uncertainty-driven selection of structures that most improve band-structure prediction. Overall, our results suggest that analytic LL-BNN active learning offers a simple yet powerful tool for data-efficient surrogate modeling. By combining principled uncertainty estimation with a lightweight implementation, the method opens the door to faster exploration and optimization of complex design spaces in photonics and other scientific disciplines.

\section{\label{sec:level1}Appendix}
\subsection{\label{sec:level2}Information-theoretic active learning}

Given model parameters $\theta$ (in our case $\mu_w, \sigma_w, \mu_b, \sigma_b$) and a labeled dataset $D$, we aim to select a new sample $(x, y)$ such that the updated dataset $D'$ minimizes the posterior entropy over the model parameters:
\begin{equation}
    \argmin_{(x,y)} H(\theta \,|\, D').
\end{equation}

Since the true label $y$ is unknown, this objective can be reformulated as its expected counterpart:
\begin{equation}
    \argmax_{x} \, H(\theta \,|\, D) - \mathbb{E}_{y \sim p(y|x, D)} \!\left[ H(\theta \,|\, y, x, D) \right],
\end{equation}
which is equivalent to maximizing the conditional mutual information $I(\theta; y \,|\, x, D)$. This expression can also be rewritten as:
\begin{equation}
    \argmax_{x} \, H(y \,|\, x, D) - \mathbb{E}_{\theta \sim p(\theta|D)} \!\left[ H(y \,|\, x, \theta) \right].
\end{equation}

In our setting, the second term vanishes because each sampled model yields a deterministic prediction (i.e., $H(y|x,\theta) = 0$). Thus, the acquisition criterion simplifies to selecting the sample that maximizes predictive entropy:
\begin{equation}
    \argmax_{x} \, H(y \,|\, x, D).
\end{equation}

For an approximate Bayesian neural network with a Gaussian last layer, the predictive distribution is given by
\[
    y = w^\top x + b \sim \mathcal{N}\!\left(\mu_{w,i} x_i + \mu_b,\; \sigma_{w,i}^2 x_i^2 + \sigma_b^2 \right).
\]
The corresponding analytic expression for the predictive entropy is then
\begin{align}
    H(y \,|\, x, D) &= \tfrac{1}{2} \log \!\left[ 2 \pi e \left( \sigma_{w,i}^2 x_i^2 + \sigma_b^2 \right) \right] \nonumber \\
    &= \tfrac{1}{2} \log \!\left[ 2 \pi e \cdot s(x) \right],
\end{align}
where $s(x)$ denotes the predictive variance. Consequently, the greedy active learning strategy reduces to selecting the sample with the highest predictive uncertainty. We note that the predictive uncertainty in this work reflects epistemic uncertainty arising from limited training data, rather than observation noise. The band gap labels are deterministically computed using the plane wave expansion solver, and we therefore assume noiseless targets. Additionally, uncertainty is modeled only in the final Bayesian linear layer, while the feature extractor remains deterministic. Consequently, the predictive variance corresponds to last-layer epistemic uncertainty conditioned on fixed learned features.

\subsection{\label{sec:level3}Per-sample uncertainty correlation}
\begin{figure}[htbp]
\centering\includegraphics[width=7cm]{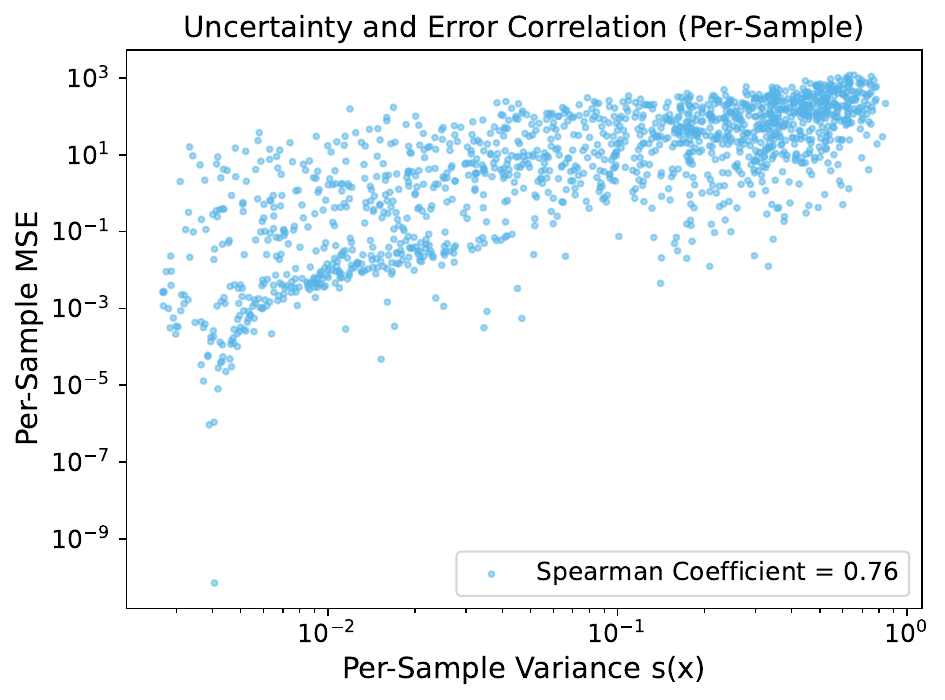}
\caption{\label{correlation_scatter_example} \textbf{Per-Sample BNN Uncertainty Correlation.} Per-sample predictive variance $s(x)$ plotted against per-sample squared error $(y-\hat{y})^2$ on a log-log scale for visualization. The positive trend confirms that higher uncertainty corresponds to higher error at the individual sample level, consistent with a Spearman coefficient of 0.76. Axes are log-scaled for clarity; no functional form is assumed.}
\end{figure}

Figure \ref{correlation_scatter_example} provides a per-sample visualization of the uncertainty-error correlation, complementing the binned calibration curve of Figure \ref{correlation_plot}. Each point represents a single test sample, with per-sample predictive variance $s(x)$ plotted against per-sample squared error $(y-\hat{y})^2$.

\subsection{\label{sec:level4}Batch redundancy analysis}

\begin{figure}[!htbp]
\centering\includegraphics[width=7cm]{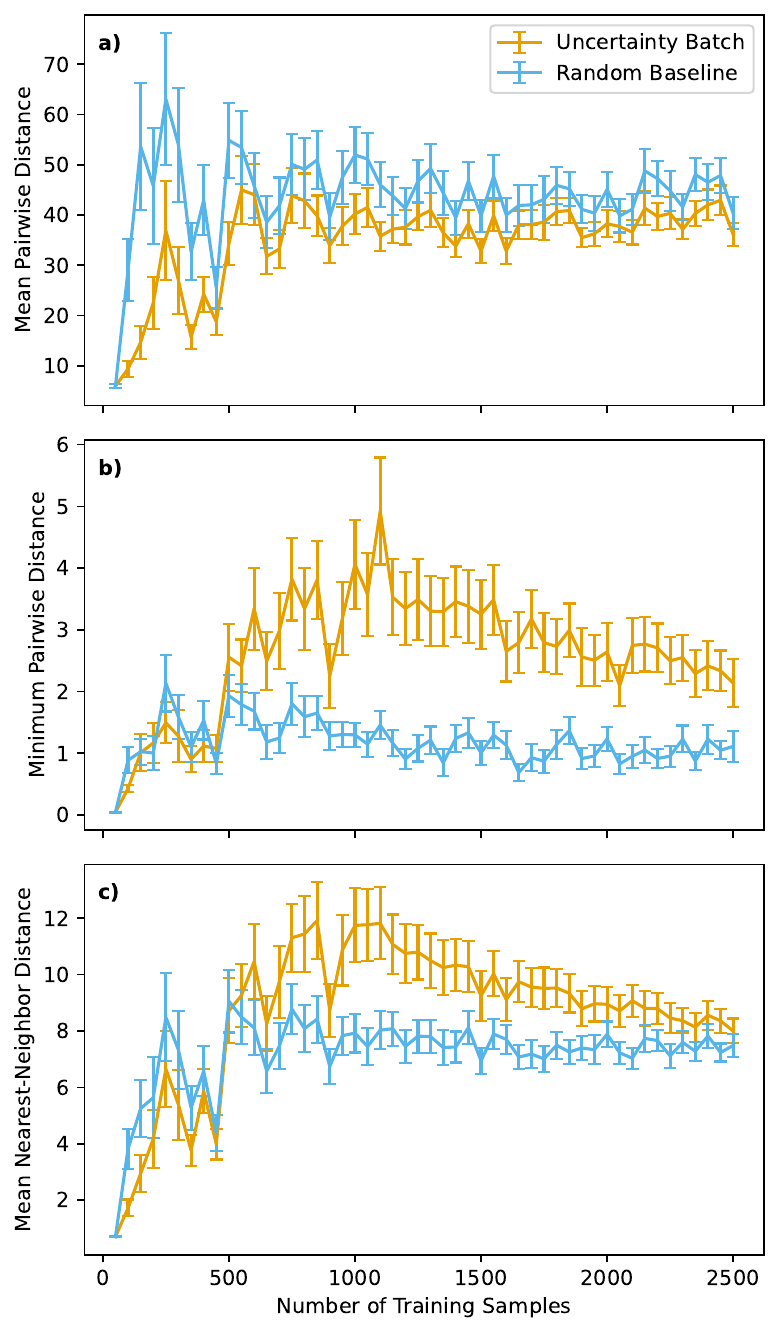}
\caption{\label{aggregated_redundancy} \textbf{Batch diversity analysis across active learning iterations.} Pairwise distances between samples selected by analytic LL-BNN uncertainty in the 128-dimensional feature space, aggregated across all 10 seeds (mean $\pm$ standard error). Orange shows the LL-BNN uncertainty-selected batch; blue shows batches from the corresponding random sampling runs, with distances computed in each run’s learned feature space. (a) Mean pairwise distance. (b) Minimum pairwise distance. (c) Mean nearest-neighbor distance. The LL-BNN batches exhibit slightly lower mean pairwise distances than random sampling, indicating mild clustering, but consistently higher minimum pairwise and mean nearest-neighbor distances, suggesting the batches do not collapse to near-duplicate selections.}
\end{figure}

To quantify batch redundancy, we measure the diversity of each selected batch in feature space at every active learning iteration. We extract the 128-dimensional backbone features of the 50 newly acquired samples and compute three Euclidean distance metrics: mean pairwise distance, minimum pairwise distance, and mean nearest-neighbor distance. Metrics are computed for both the analytic LL-BNN and random sampling within each run’s learned feature space, aggregated across 10 seeds, and reported as mean ± standard error.
Because absolute distances in high-dimensional feature spaces are difficult to interpret, we compare results relative to the random baseline. As shown in Figure \ref{aggregated_redundancy}, the analytic LL-BNN yields slightly lower mean pairwise distances but consistently higher minimum pairwise and mean nearest-neighbor distances, indicating mild clustering without near-duplicate selections.

\pagebreak

\noindent
\begin{minipage}[t]{\columnwidth}
\textbf{Funding.}
We would like to thank the Dean of Science Fellowship for funding this study. This research was also sponsored in part by the United States Air Force Research Laboratory and the Department of the Air Force Artificial Intelligence Accelerator and was accomplished under Cooperative Agreement number FA8750-192-1000. The views and conclusions contained in this document are those of the authors and should not be interpreted as representing the official policies, either expressed or implied, of the Department of the Air Force or the US Government. The US Government is authorized to reproduce and distribute reprints for Government purposes notwithstanding any copyright notation herein. This material is also based upon work sponsored in part by the US Army DEVCOM ARL Army Research Office through the MIT Institute for Soldier Nanotechnologies under Cooperative Agreement number W911NF-23-2-0121, as well as MIT MGAIC and Shell Global. This work is also supported by the National Science Foundation under Cooperative Agreement PHY-2019786 (The NSF AI Institute for Artificial Intelligence and Fundamental Interactions, \href{http://iaifi.org/}{\texttt{http://iaifi.org/}}).

\medskip
\noindent\textbf{Acknowledgment.}
We would like to thank MIT SuperCloud for computing resources and acknowledge ChatGPT for assistance with manuscript text editing.

\medskip
\noindent\textbf{Disclosures.}
The authors declare no conflicts of interest.

\medskip
\noindent\textbf{Data Availability Statement.}
Data and code underlying the results presented in this paper are available at
\href{https://github.com/ryanlopezzzz/photonics_al}{
\texttt{https://github.com/ryanlopezzzz/photonics\_al}}.
\end{minipage}


\FloatBarrier
\bibliographystyle{apsrev4-2-longbib}
\bibliography{refz-manual-fixes}

\makeatletter\@input{createsiaux.tex}\makeatother
\end{document}